\documentclass[
twocolumn,preprintnumbers
]{revtex4}
\usepackage{amssymb}
\usepackage{graphicx}
\usepackage{dcolumn}
\usepackage{bm}
\usepackage{epsfig}
\usepackage{amssymb}
\usepackage{color}

\newcommand{\bal}{\begin{align}}
\newcommand{\eal}{\end{align}}
\newcommand{\beq}{\begin{eqnarray}}
\newcommand{\eeq}{\end{eqnarray}}
\newcommand{\nneeq}{\nonumber \end{eqnarray}}
\newcommand{\nnn}{\nonumber}
\newcommand{\nn}{\nonumber \\}
\newcommand{\es}{& = &}
\newcommand{\rs}{\, = \,}
\newcommand{\ps}{& + &}
\newcommand{\ms}{& - &}
\newcommand{\ts}{& \times &}
\newcommand{\nt}{\nn \ts}
\newcommand{\np}{\nn \ps}
\newcommand{\nm}{\nn \ms}

\newcommand{ \h}{ \,{ 1 \over 2 }\, }

\newcommand{\cM}{ {\cal M} }

\newcommand{\cT}{ {\cal T} }

\begin{document}
\title{ The Ridge Effect, Azimuthal Correlations, and other Novel Features of Gluonic String Collisions in High Energy Photon-Mediated Reactions }
\author{ Stanis{\l}aw D. G{\l}azek }
\email{stglazek@fuw.edu.pl}
\affiliation{ Institute of Theoretical Physics\\ 
Faculty of Physics, University of Warsaw, Pasteura 5, 02-093 Warsaw, Poland }
\author{ Stanley J. Brodsky }
\email{sjbth@slac.stanford.edu}
\affiliation{SLAC National Accelerator Laboratory\\
Stanford University, Stanford, California 94309, USA}
\author{ Alfred S. Goldhaber }
\email{goldhab@max2.physics.sunysb.edu}
\affiliation{ C.N. Yang Institute for Theoretical Physics\\
 State University of New York, Stony Brook, NY 11794-3840, USA}
\author{ Robert W. Brown }
\email{rwb@case.edu }
\affiliation{Department of Physics\\ Case Western Reserve University\\
Rockefeller Bldg., 2076 Adelbert Rd., Cleveland, OH 44106, USA}
\date{ May 21, 2018 }

\begin{abstract}
One of the remarkable features of high-multiplicity hadronic
events in proton-proton collisions at the LHC is the fact
that the produced particles appear as two ``ridges",
opposite in azimuthal angle $\phi$, with approximately 
flat rapidity distributions. This phenomena can be identified 
with the inelastic collision of gluonic flux tubes associated 
with the QCD interactions responsible for quark confinement 
in hadrons.  In this paper we analyze the ridge phenomena 
when the collision involves a flux tube connecting the quark 
and antiquark of a high energy real or virtual photon. We 
discuss gluonic tube string collisions in the context of two 
examples: electron-proton scattering at a future electron-ion 
collider or the peripheral scattering of protons accessible at 
the LHC. A striking prediction of our analysis is that the 
azimuthal angle of the produced ridges will  be correlated 
with the scattering plane of the electron or proton producing 
the virtual photon. In the case of $ep \to eX$, the final state 
$X$ is expected to exhibit maximal multiplicity when the elliptic 
flow in $X$ is aligned with the electron scattering plane. In 
the $pp \to ppX$ example, the multiplicity and elliptic flow in 
$X$ are estimated to exhibit correlated oscillations as functions 
of the azimuthal angle $\Phi$ between the proton scattering 
planes. In the minimum-bias event samples, the amplitude of 
oscillations is expected to be on the order of 2\% to 4\% of 
the mean values. In the events with highest multiplicity, the 
oscillations can be three times larger than in the minimum-bias 
event samples.
\end{abstract} 

\preprint{SLAC-PUB-17260}

\maketitle                 


\section{ Introduction }
\label{intro}

Scattering has been critical to understanding of submicroscopic 
structure since the earliest stages.  Alpha-particle scattering 
revealed the atomic nucleus, and its components -- protons 
and neutrons.  Later, elastic electron scattering found the electric 
charge distribution of nuclei, and inelastic electron scattering found 
point-like structure inside the nucleon, while electron-positron pair 
annihilation revealed also the presence of gluons inside hadrons. 
The existence of such objects does not necessarily tell the whole 
story about hadron structure: the arrangement of quark and gluon 
configurations could be more than just `floating' particles. 
One possible configuration is a string of gluons connecting quark 
and anti-quark in a meson, or quark and di-quark in a nucleon. 
Strings provide a linear structure, so that if two strings are aligned 
parallel to each other in the plane perpendicular to the collision 
direction, then their collision can give rise to higher multiplicity 
than if the strings are oriented transversely to each other. The 
flow of collision products is also different. It is hence encouraging 
for thinking about  gluon strings that patterns of varying 
multiplicity and collective flow of products are seen in occasional 
$pp$ collisions. 

\subsection{ Strings in $ep \to eX$ and $pp \to ppX$ scattering }

\begin{figure}[ht!]
          \includegraphics[width=0.25\textwidth]{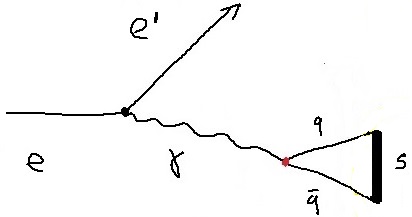}
          \caption{ Scattering of an electron yields a virtual photon 
          that develops quark-anti-quark pair connected by a gluon 
          string.
          \label{fig:string}}
          \end{figure}

\begin{figure}[ht!]
          \includegraphics[width=0.3\textwidth]{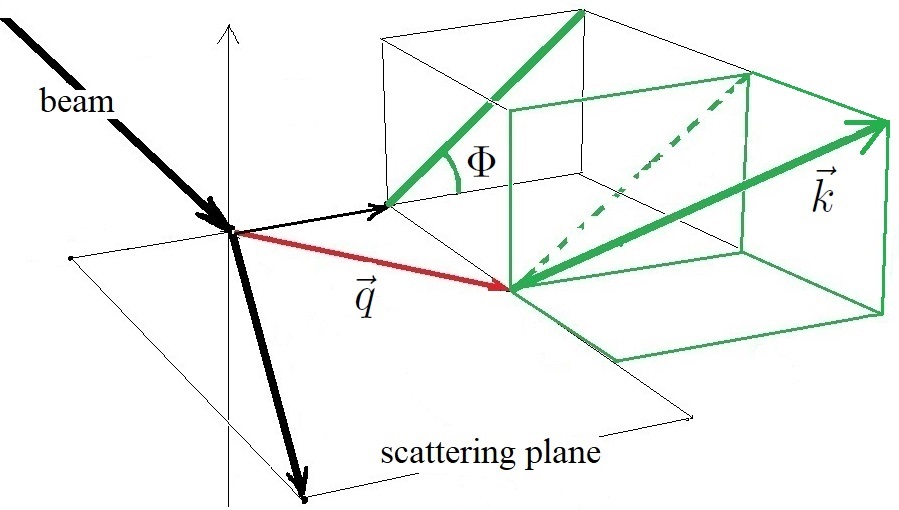}
          \caption{ Beam of charged particles is scattered with momentum 
          transfer $\vec q$ that defines a scattering plane. The photon that 
          carries $\vec q$, creates a quark-anti-quark pair with relative momentum
          $\vec k$. The gluon string lies along that momentum and it forms 
          the azimuthal angle $\Phi$ with the scattering plane. 
          \label{fig:Phi}}
          \end{figure}

\begin{figure}[ht!]
          \includegraphics[width=0.4\textwidth]{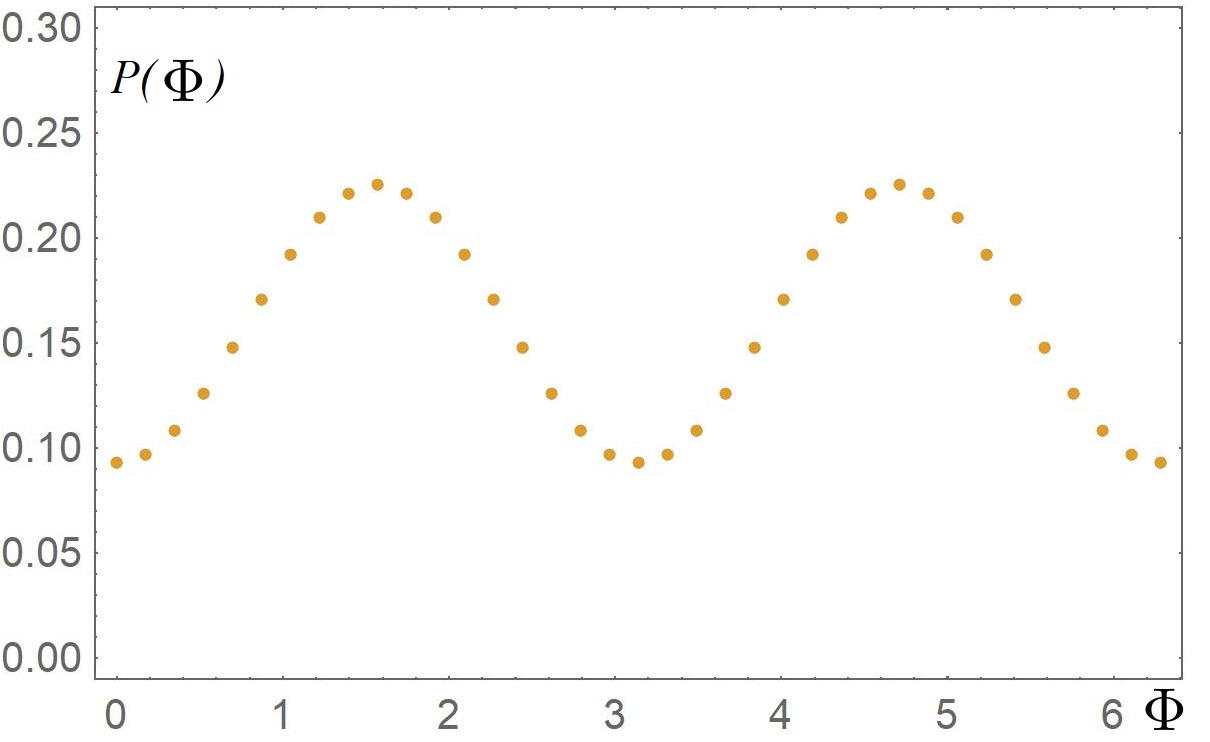}
          \caption{ Example of probability $P(\Phi)$ of finding a gluon string produced 
          electromagnetically through scattering of an electron, as a function of the 
          azimuthal angle $\Phi$ around the electron beam between the electron 
          scattering plane and the string axis that connects quark with anti-quark in 
          the string rest frame, see Fig.~\ref{fig:Phi}. In the example, the incoming electron momentum is 
          7 TeV. Electrons  scatter elastically with momentum transfers of square 
          $q^2 = -1.4$~GeV$^2$, gaining 1 GeV of momentum transverse to the 
          beam and losing 1/4 of their initial momentum along the beam. Only the 
          strings due to $u \bar u$ are accounted for. Inclusion of $d$ and $s$ quark 
          pairs reduces the probability variation amplitude by about one percent. 
          Inclusion of heavy quarks further reduces the variation by about one fifth 
          of the shown magnitude, mostly due to the $c$ quarks that cause about 
          98\% of the latter reduction. The probability plot practically does not change 
          when the electron beam is replaced by a proton beam.
          \label{fig:PPhi}}
          \end{figure}

We suggest that scattering of charged particles may be directly 
sensitive to the formation of gluon strings~\cite{Photon2017}. 
An electron can emit a virtual photon that in turn develops a 
quark-anti-quark pair connected by a gluon string (see 
Fig.~\ref{fig:string}). 
The string azimuthal orientation, defined by the angle $\Phi$ illustrated in Fig.~\ref{fig:Phi}, is correlated with the scattering plane of the parent charged particle.
 An example of the probability distribution of string orientation with respect to the scattering
 plane is shown in Fig.~\ref{fig:PPhi}. Details of its calculation are described in 
 Sec.~\ref{elements}.
Such a  probability distribution implies that the charged-particle 
scattering-plane azimuthal orientation is correlated with the 
multiplicity or collective flow of particles that emerge from a
subsequent collision of the virtual string with another one that 
comes from the opposite-going charged particle in the scattering 
process. 

\begin{figure}[ht!]
\includegraphics[width=0.45\textwidth]{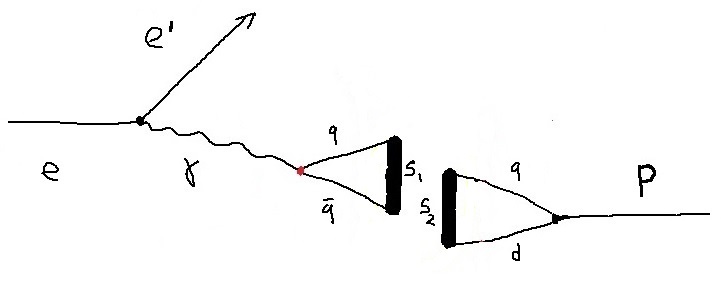}
\caption{ Electron scattering on protons proceeds through
             a collision of gluon strings, $S_1$ that connects a quark
             and anti-quark in a photon and $S_2$ that connects 
             a quark and a di-quark in a proton. } 
         \label{fig:ep} 
\end{figure}

\begin{figure}[ht!]
\includegraphics[width=0.45\textwidth]{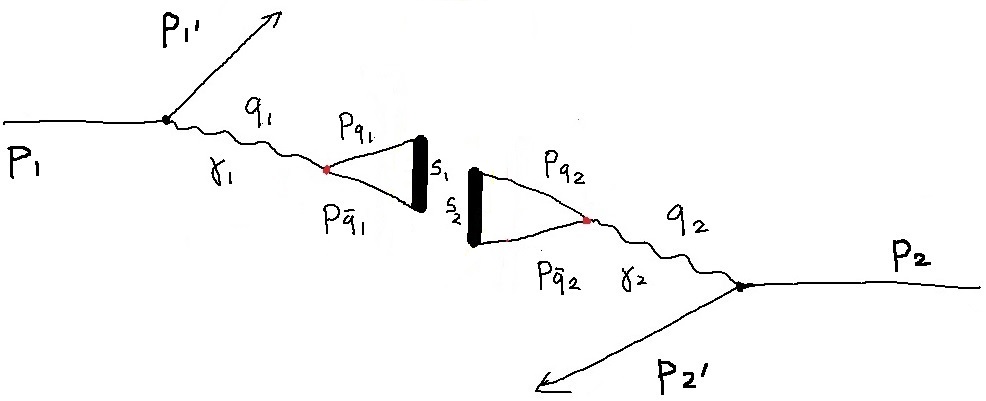}
\caption{ The peripheral $p_1p_2 \to p_{1'} p_{2'} X$ scattering 
              proceeds through collision of gluon strings $S_1$ and 
              $S_2$.} 
         \label{fig:pp} 
\end{figure}

\begin{figure}[ht!]
          \includegraphics[width=0.4\textwidth]{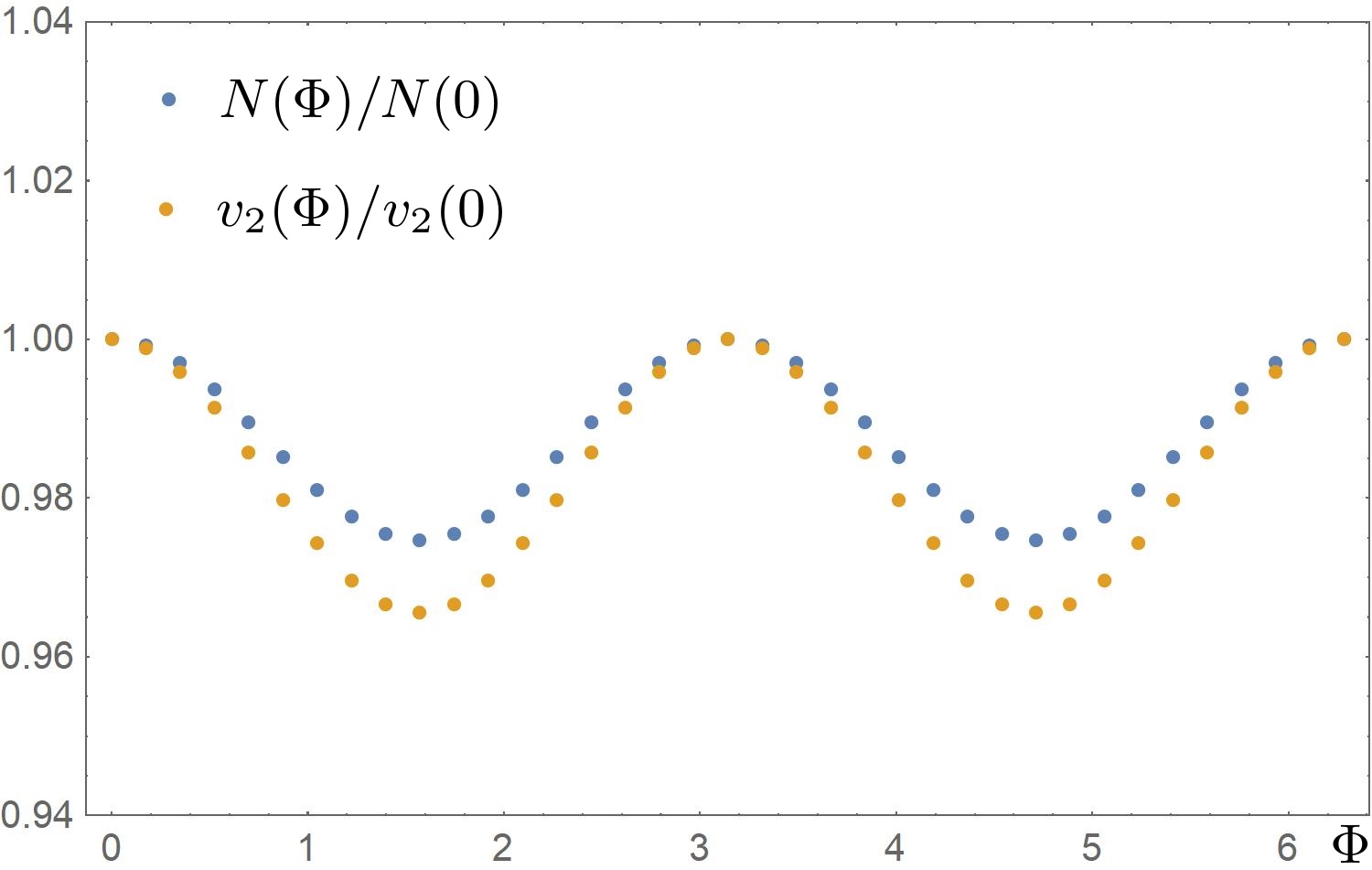}
          \caption{ Example of an estimate for the minimal-bias average ratios 
          of multiplicity, $N(\Phi)/N(0)$, and elliptic flow, $v_2(\Phi)/v_2(0)$,
          in the $pp$ peripheral scattering process $p_1p_2 \to p'_1 p'_2 X$ 
          at LHC that proceeds through collisions of gluon strings. The 
          ratios are calculated using Eqs.~(\ref{NMB}) and (\ref{epsMB}). 
          They depend on the angle $\Phi$ between the two projectile 
          planes that are shown in Fig.~\ref{fig:AnglePhi}. Protons come 
          from opposite directions, with momenta equal in size, 7 TeV. 
          They scatter elastically with momentum transfers 
          squared $q_1^2 = q_2^2 = -1.4$~GeV$^2$. For example, proton 
          $p_1$ scatters horizontally, gaining 1 GeV of momentum transverse 
          to the beam and losing 1/4 of its initial momentum along the beam. 
          Proton $p_2$ scatters in a plane forming angle $\Phi$ with the 
          horizontal plane and gains 1 GeV of momentum in that plane 
          transversely to the beam and also loses 1/4 of its initial momentum 
          in the beam direction. The multiplicity is estimated assuming it is 
          proportional to the number of partonic collisions and the elliptic 
          flow is estimated assuming it is proportional to the eccentricity in 
          the initial stage of the string collision. The unknown proportionality 
          constants cancel out in the shown ratios. 
          \label{fig:example}}
          \end{figure}

\begin{figure}[ht!]
          \includegraphics[width=0.3\textwidth]{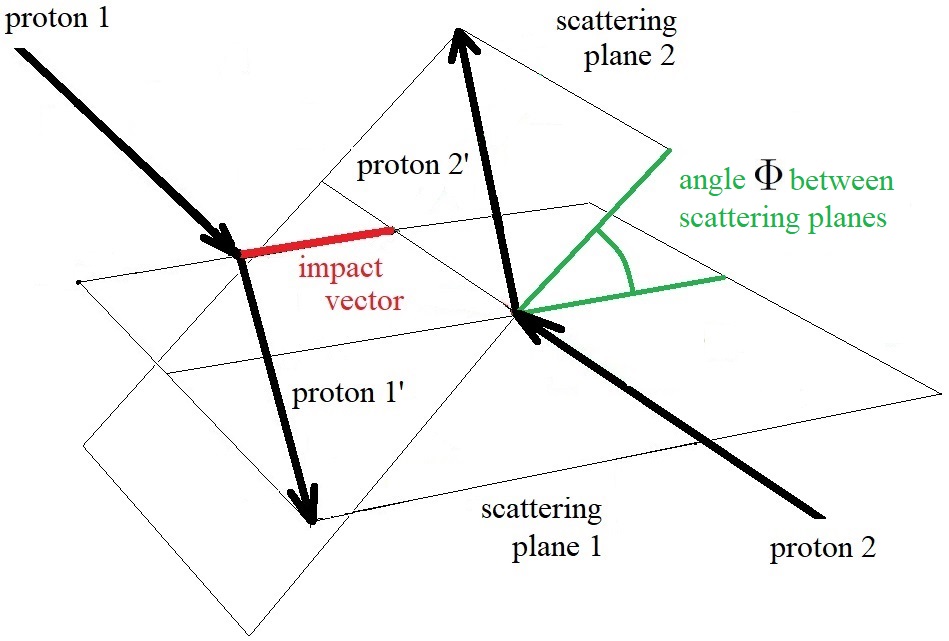}
          \caption{ Angle $\Phi$ between scattering planes in Fig.~\ref{fig:example}.
          \label{fig:AnglePhi}}
          \end{figure}

To be specific, when an electron scatters off a proton, through 
a collision of a string of gluons in a photon with a string that 
connects a quark and a di-quark in the proton, see Fig.~\ref{fig:ep}, 
the multiplicity of the final state is optimized when its collective 
flow is aligned with the electron scattering plane. Similarly, the 
probability of rare events with large multiplicity and elliptic flow 
in peripheral $pp$ scattering illustrated in Fig.~\ref{fig:pp}, 
{\it cf.}~\cite{Bjorken:2013boa,Lithuania,Venugopalan1,Venugopalan2}, 
is maximal when one projectile recoil is parallel to the recoil of 
the opposite-going projectile. For example, such rare 
large-multiplicity peripheral events with sizable elliptic flow 
at LHC are expected to have maximal probability of occurring 
when the two proton scattering planes are parallel to each other. 
The probability is minimal when the scattering planes of the protons 
appear rotated by 90 degrees with respect to each other. 
Figure~\ref{fig:example}
illustrates our estimate for string 
collision effects in terms of the multiplicity $N(\Phi)/N(0)$ 
and elliptic flow $v_2(\Phi)/v_2(0)$ in the final state $X$ as 
functions of the angle $\Phi$ between the planes $(p_1, p'_1)$ 
and $(p_2, p'_2)$ defined by the direction of proton beams 
and final proton three-momenta $\vec p\,'_1$ and $\vec  p\,'_2$ 
in the laboratory, see Fig.~\ref{fig:AnglePhi}.
Figure~\ref{fig:example} is obtained using Eqs.~(\ref{NMB}) and 
(\ref{epsMB}). Examples like this suggest that the azimuthal 
variation of multiplicity and elliptic flow can be used to study 
properties of gluonic strings in LHC.

\begin{figure}[ht!]
          \includegraphics[width=0.25\textwidth]{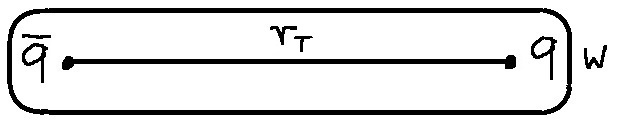}
          \caption{ Qualitative view of the two-dimensional 
          string shape in the transverse plane (TP). 
          The length $r_T$ corresponds to the vector $\vec r_T$ 
          that describes the relative position of the quark with 
          respect to the anti-quark, projected on the TP. 
          $w$ denotes the string width. The shape is assigned 
          some thickness distribution, $\rho(\vec x_T)$, that 
          corresponds to the density of partons as function of 
          position $\vec x_T$ in the TP. The shape of string 
          ends does not matter much if the string length is much 
          bigger than width. In this work, the ends are rectangular, 
          which is the simplest shape to use in our estimates.         
          \label{fig:shape}}
          \end{figure}

\begin{figure}[ht!]
          \includegraphics[width=0.15\textwidth]{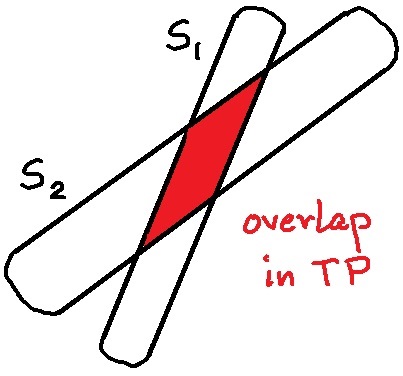}
          \caption{  The number of partonic collisions in a collision
          of two strings $S_1$ and $S_2$, is assumed to be 
          proportional to the overlap area in the TP, in which
          the strings are seen as they appear when looked
          at along the beam.
         \label{fig:overlap}}
         \end{figure}

Our calculations are carried out using  Hamiltonian dynamics 
in quantum field theory in the approximation of very large beam 
momentum, and collisions of strings are estimated using a 
geometrical picture. The strings seen along the proton beam 
form certain shapes (see Fig.~\ref{fig:shape}) 
in the plane transverse to the beam. Below, this plane is called  
the transverse plane (TP). The string shape in the TP corresponds
to a string that is stretched in space along the vector $\vec r$, 
which extends from the anti-quark to quark in the string rest 
frame (SRF). When the string moves very fast along the 
proton beam, its shape in the TP is seen in the laboratory 
as built around a two-dimensional vector $\vec r_T$ that 
forms the transverse part of $\vec r = (r_x, \vec r_T)$. 
The component $r_x$ corresponds to the beam direction 
in the frame of reference that we work with; its $x$-axis 
is chosen along the beam. The collision of strings $S_1$ 
and $S_2$ proceeds via interaction of partons in the region 
of overlap of the string shapes on the TP, illustrated in 
Fig.~\ref{fig:overlap}. 
         
The magnitude of overlap grows when the angle between the 
strings decreases, provided that their impact parameter is 
smaller than the string width. There is a competing effect 
that the impact parameters may reach values comparable
with the string length, much larger than the string width,  
when the angle between the strings is not small. Therefore,
in the minimal bias (MB) averages over events with all possible 
impact parameters, these effects cancel each other. The 
resulting azimuthal correlations between projectile planes
and the average multiplicity and elliptic flow, are small. They 
can be enhanced by averaging over events with relatively 
large multiplicities, discussed later.
 
Concerning the length and width of gluonic strings, we 
assume that the typical time needed by a string to reach 
its full width is longer than the time of stretching a string 
by quarks. This is in agreement with the flux-tube knot 
model assumption~\cite{StringKnots1,StringKnots2} that 
the relaxation of a topologically non-trivial string configuration 
to a tight-knot state configuration is faster than the configuration 
decay rate, {\it cf.}~\cite{JohnsonThorn,IsgurPaton}. In 
the same spirit, we assume that the suddenly stretched 
string has a diameter as small as $\sim 0.1$ fm. 
Knowing that hadrons extend over distances order 1 fm, 
one could consider string models with width related to 
length, such as $w = r/10$ etc.

One might hope to estimate the shape of gluonic strings 
in photons using the vector dominance model (VDM) of 
photon-nucleon coupling. Three issues arise.  One is that 
the VDM does not provide any information on the shapes 
of flux tubes in virtual $\rho$ mesons. The other issue is 
that a collision of two photons may proceed through 
quark-anti-quark pairs whose relative motion is not at all 
limited as the relative motion of quarks is limited by the
wave function of quarks in the neutral $\rho$-meson. 
The quark-gluon picture is dual to the sum over the whole 
spectrum of hadronic states that contribute. The third issue 
is that although the VDM suggests that a photon can interact 
with an extended proton via a pair of quarks resembling a 
$\rho$-meson, it does not tell us how a photon can turn 
into a pair of quarks that do not interact and do not 
need to overlap with a nucleon's structure. Having no firm 
theoretical input concerning the relation between strings' 
width $w$ and the distance between quark and anti-quark 
in a pair in a photon, $r = |\vec r \,|$, we consider $w$ a 
free parameter. Figure~\ref{fig:example} is obtained 
assuming that $w = 0.1$ fm.

The magnitude of string overlap is used as a measure of 
the number of partonic collisions that are possible in a 
given configuration of strings heading towards each other. 
Multiplicity is assumed proportional to this number. The 
shape of the overlap area is used to estimate the eccentricity 
of the colliding parton-matter in the transverse plane. The 
eccentricity is used to estimate the elliptic flow in $X$, the 
latter assumed proportional to the former. The unknown 
proportionality constants cancel out in ratios shown in 
Fig.~\ref{fig:example}. 
   
In an event that leads from protons $p_1$ and $p_2$ 
to $p'_1$ and $p'_2$, the density of parton-parton 
collisions in the TP is described using the 
formula~\cite{Glauber,v2varepsilon2,dentPLB,blaizot}.
\beq
\label{ncoll}
n_{\rm coll}(\vec x_T, \vec b , \vec r_1, \vec r_2 ) 
\es
\sigma 
\,  
\rho_1(\vec x_T - \vec b/2, \vec r_1)
\,
\rho_2(\vec x_T + \vec b/2, \vec r_2) . \nn
\eeq
It says that the probability density for finding 
pairs of partons capable of colliding at the point 
$\vec x_T$ in the TP, is proportional to a product 
of parton probability densities in the parent strings 
that are separated by the impact vector $\vec b$. 
The photons come from protons almost along the beam 
and the string impact parameter is assumed the same as 
the proton-proton impact parameter. The coefficient
$\sigma$ stands for the parton-parton cross-section,
presumably on the order of a few to ten mb. The string 
parton densities are determined by the string orientations, 
$\vec r_1$ and $\vec r_2$, in the respective SRFs. Formulas 
used for estimates of the parton densities in strings are 
described in Sec.~\ref{geometricaloverlap}. 

In addition to the string parton densities, evaluation
of observables concerning collisions of strings involves 
probabilities, denoted below by $P(\vec r_1, p'_1,p_1)$ 
and $P(\vec r_2, p'_2,p_2)$, for the string orientation 
vectors $\vec r_1$ and $\vec r_2$. These probabilities 
are estimated using QED. They depend on the incoming 
and outgoing protons'  momenta. Protons' spins are 
summed and averaged over because we assume that 
the beam protons are not polarized and final proton 
polarizations are not measured. Strings are assumed 
to not depend on the quark spins. The probabilities
we use are described in Sec.~\ref{Pstring}.

\subsection{ Observables associated with string collisions }

Consider the example of $pp$ scattering. The density 
of collisions in Eq.~(\ref{ncoll}) is used to estimate 
multiplicity and ridge effects as functions of the 
azimuthal angle $\Phi$ between proton planes. The 
multiplicity $N(\vec b, \vec r_1, \vec r_2)$ of $X$ is 
assumed proportional to the number of partonic collisions, 
\beq
N(\vec b, \vec r_1, \vec r_2)
\es 
C_N \ N_{\rm coll}(\vec b, \vec r_1, \vec r_2) \\
\es
C_N \ \int d^2 x_T \
n_{\rm coll}(\vec x_T, \vec b , \vec r_1, \vec r_2) \ .
\eeq
$C_N$ denotes an unknown coefficient that
can be estimated by comparison with models of string
collisions and data on particle production for selected 
scattering parameters.

The ridge effect is estimated assuming that the elliptic 
flow $v_2$ in $X$ is proportional to the eccentricity 
$\varepsilon_2(\vec b, \vec r_1, \vec r_2)$ of the density 
of partonic collisions, $n_{\rm coll}(\vec x_T, \vec b , 
\vec r_1, \vec r_2 )$, with a model-dependent coefficient 
on the order of 0.3~\cite{v2varepsilon2}. Definition of 
eccentricity~\cite{blaizot} uses the concept of averaging 
of a quantity $f(\vec x_T)$ with the density,
\beq
\{ f \} 
\es 
\frac{\int d^2 x_T \ f(\vec x_T) \ n_{\rm coll}(\vec x_T, \vec b , \vec r_1, \vec r_2 )  }
{\int d^2x_T \ n_{\rm coll}(\vec x_T, \vec b , \vec r_1, \vec r_2) } \ ,
\eeq
where it is understood that $ \vec x_T $ is measured from
the geometrical center of the overlap area in the TP so that
$\{ \vec x_T \}=0$. One introduces polar coordinates in the
TP, angle $\alpha$ and length $x_T = |\vec x_T|$, using 
$\vec x_T = (x,y) = x_T \ (\cos{\alpha},\sin{\alpha})$. 
Eccentricity is defined in terms of the averaged values 
of $x^2 - y^2 = x_T^2 \cos 2\alpha$ and $2xy = x_T^2 
\sin 2 \alpha$. Both are contained in the averaged value 
of the complex quantity $x_T^2 e^{2 i \alpha}$. Eccentricity 
of a density is defined as the modulus of averaged value
of this quantity~\cite{blaizot}
\beq
\label{epsilonn+1}
\varepsilon_2(\vec b, \vec r_1, \vec r_2) 
\es
\frac{\sqrt{\{ x_T^2 \cos(2 \alpha) \}^2 
+\{ x_T^2 \sin(2 \alpha)\}^2}}{\{ x_T^2 \} } \ .
\eeq
The square root of sum of squares of averaged values 
reflects the relationship between real and imaginary 
parts of a complex number and its modulus.

The differential string-collision cross section can be estimated using 
the formula~\cite{Glauber,dentPLB,Goldhaber1,Goldhaber2}
\beq
\label{dsigma}
{d^2 \sigma (\vec b, \vec r_1, \vec r_2 \, )\over  db_y db_z } 
\es
1 -  e^{- N_{\rm coll}(\vec b, \vec r_1, \vec r_2) } \ ,
\eeq
which corresponds to the assumption that the dominant 
angle dependence comes from strings that are made 
of  similar numbers of gluons, $N_g$, large enough to 
justify the use of an exponential form. Otherwise, a formula $(1 - 
N_{\rm coll}/N_g^2)^{N_g^2}$ may be used. The 
examples shown in Fig.~\ref{fig:example} are obtained
using Eq.~(\ref{dsigma}). 

Estimates for the MB-averaged $\Phi$-dependence of multiplicity 
and elliptic flow, shown in Fig.~\ref{fig:example}, result 
from integration over the impact vector $\vec b$. We use 
the formulas
\beq
\label{NMB}
N(p'_1,p'_2) 
\es 
{C_N \over \sigma_{\rm p } }
\int d^2b 
\nt
\int d^3r_1 \ P(\vec r_1, p'_1,p_1)
\int d^3r_2 \ P(\vec r_2, p'_2,p_2)
\nt
\left[ 1 - e^{-N_{\rm coll}(\vec b, \vec r_1, \vec r_2)} \right]  
\ N_{\rm coll} (\vec b, \vec r_1, \vec r_2) \ , \\
\label{epsMB}
v_2(p'_1,p'_2)
\es
{C_v \over  \sigma_{\rm p } }
\int d^2b 
\nt
\int d^3r_1 \ P(\vec r_1, p'_1,p_1)
\int d^3r_2 \ P(\vec r_2, p'_2,p_2)
\nt
 \left[ 1 - e^{-N_{\rm coll} (\vec b, \vec r_1, \vec r_2)} )\right] 
 \ \varepsilon_2(\vec b, \vec r_1, \vec r_2) \ ,
\eeq
where  $\sigma_{\rm p }$ is a total cross section normalization 
factor that cancels out in the ratios of interest in Fig.~\ref{fig:example}.
The normalization factor will not be further discussed. We only mention 
that its calculation requires integration over all possible final states. 
Similarly, the unknown proportionality constant $C_v$, assumed to 
relate elliptic flow to eccentricity, cancels out in the ratios of 
Fig.~\ref{fig:example}. Consequently, we focus on the eccentricity
that is meant to yield directly the ratio shown in Fig.~\ref{fig:example} 
for the elliptic flow.

\section{ Geometrical overlap of strings }
\label{geometricaloverlap}

Modeling of string collisions involves consideration of rotations 
and boosts required for description of the off-shell string 
world-sheets in different frames of reference. However, the 
fast motion of a quark-di-quark or quark-anti-quark pair 
along the beam implies a simple picture in the TP. We
limit explicit discussion to quark-anti-quark pairs from
photons. 

\subsection{ String shape in the  transverse plane }

In the laboratory, a string specified by the quark-anti-quark 
relative position vector, $\vec r$ in the SRF, moves very fast 
along the proton beam. The beam is used to define the 
laboratory $x$-axis. The string has the same width $w$, 
transverse to the beam, in both frames. The length and orientation 
of the string on the TP are determined by the vector $\vec r_T$, 
which is the transverse part of $\vec r$, and a correction due 
to the string width (see below). The string shape in the TP has 
some density profile, coming from the string structure and 
orientation in space. In the TP region where the string density 
differs from zero, it is useful to set the string thickness to its 
average value, say, $\bar \rho_1$ for string $S_1$ and
 $\bar \rho_2$ for $S_2$. This approximation greatly simplifies 
 estimates of angular correlations. 

If a string forms an angle $\beta$ with the TP, which is the
angle it makes with the $yz$-plane in the SRF, its projection 
on that plane is described by the vector $\vec l$ of length  
\beq
l \es (r + w) |\cos{\beta}| + w |\sin{\beta}|  \  , 
\eeq
 where $r = |\vec r \, |$ 
and $\sin \beta = r_x/ r$. The two-dimensional string shape 
in the TP is assumed to be a rectangle of  area $a_\beta 
= w l$.  The string average parton probability density on 
the TP, $\bar \rho$, times the area $a_\beta$, gives the same 
number of partons, $N_g$, described by the formula $\bar 
\rho a_\beta$,  as a product of the average three dimensional 
parton probability density in the string, $\rho$, times the string 
volume, $ \rho (r+w) \pi w^2/4 $. This implies
\beq
\label{rho}
\bar \rho  \es  { \rho \pi w (r+w)/4 \over (r+w) |\cos{\beta}| + w |\sin{\beta}| } \ .
\eeq
In this approximation, the product of parton densities is 
\beq
\label{overlapA}
&&
\rho_1(\vec x_T - \vec b/2, \vec r_1) \ \rho_2(\vec x_T + \vec b/2, \vec r_2) \nn
\es
\bar \rho_1 \ \bar \rho_2 \
A_1(\vec x_T - \vec b/2, \vec r_1) \ A_2(\vec x_T + \vec b/2, \vec r_2) \ ,
\eeq
where $A_1$ and $A_2$ denote the characteristic functions
of the two string shapes on the TP. The two strings 
may {\it a priori} have different widths, $w_1$ and $w_2$. These
widths may be correlated with the string lengths, as an element 
of modeling of how strings develop. The characteristic function 
of the overlap area is denoted by $A(\vec x_T, \vec b ; \vec r_1, \vec r_2) 
 = A_1(\vec x_T - \vec b/2, \vec r_1) \  A_2(\vec x_T + \vec b/2, \vec r_2) $. 
 In our estimates, we only consider fixed, small $w_1 = w_2 = w= 1/10$ fm.
 
\subsection{ Thin string approximation }

\begin{figure}[ht]
          \includegraphics[width=0.3\textwidth]{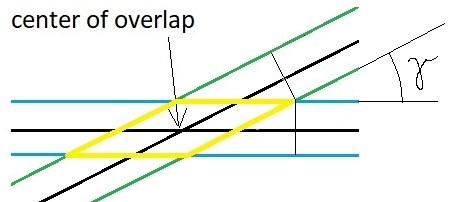}
          \caption{ Rhombus  overlap area for thin strings.
          \label{fig:intx-overlapcalculation-20171217} }
          \end{figure}

The thin strings approximation means that the string
width $w$ is typically much smaller than its length
$r$. For thin strings, the dominant overlap shape 
is a rhombus, as illustrated in Fig.~\ref{fig:intx-overlapcalculation-20171217}.
It is dominant in the sense that a rhombus occurs 
for most  of the impact parameters and for most 
lengths $l_1$ and $l_2$ of the strings and most 
angles $\gamma$ between them. The rhombus 
area and eccentricity are
\beq
\label{arhombus}
a_{\rm rhombus} \es w^2/ |\sin \gamma \, |   \  , \\
\label{erhombus}
\varepsilon_{2 \, \rm rhombus} \es |\cos \gamma \, | \ . 
\eeq
Note that these quantities do not depend on the string
lengths. Departures from the rhombus shape occur for 
small $\gamma$ or when one string overlaps the other 
with its end. 
 
\section{ Probability of a string }
\label{Pstring}

We consider $pp$ scattering. Equations (\ref{NMB}) 
and (\ref{epsMB}) provide expectation values of 
multiplicity $N(\vec b, \vec r_1, \vec r_2)$ and 
eccentricity $\epsilon(\vec b, \vec r_1, \vec r_2)$, 
averaged over string lengths and orientations described 
in terms of vectors $\vec r_1$ and $\vec r_2$. The 
probability densities, $P(\vec r_1, p'_1,p_1)$ and 
$P(\vec r_2, p'_2,p_2)$, are estimated using the canonical 
QED Hamiltonian in the instant form (IF) of 
dynamics~\cite{Dirac1} in the Coulomb, or radiation 
gauge~\cite{BD}, in which protons are treated as 
extended particles with form factors, and quarks are 
coupled to photons as point-like particles. Formulas 
for electrons are obtained by removing the form factors. 
The limit of large projectile energy allows us 
also to use the infinite momentum frame approximation,
which leads to the advantage of using results of the 
front form of dynamics and, in particular,  light-front
holography~\cite{zeta1,zeta2} for estimates of the 
string probability, in addition to the estimates based
on the IF linear potential, {\it e.g.}, see Sec.~\ref{kr}.

The quark-anti-quark pair creation term in the Hamiltonian 
acts off energy shell. Therefore, when the pair current is
contracted with the physical proton momentum transfer, 
one does not obtain zero, even if one assumes, as we do,
that the electromagnetic production of off-shell $q \bar q$ 
pairs is fully expressible in terms of the physical momentum 
transfer $q^\mu = p^\mu - p'^\mu$ between the initial 
proton $p$ and the final one, $p'$. In order to remove 
the off-shell current non-conservation effect, one can 
replace the canonical pair current off shell by
\beq
\label{jcons}
j^\mu_{\rm pair}  & \to & 
j^\mu_{\rm cons}  \rs j^\mu_{\rm pair} - { j_{\rm pair} q \ q^\mu \over q^2}  \ .
\eeq
Since the proton current is conserved, the conserved 
pair current coupling through $g_{\mu \nu}$ to the 
proton current, $j_{\rm cons} j_{\rm proton}$ is the 
same as $j_{\rm pair}j_{\rm proton}$. The same 
result is obtained using the Feynman gauge.
 
\subsection{ String amplitude }

The Hamiltonian terms that describe the dynamics of 
quark pairs with strings, $q\bar q S$, and protons, $p$, 
as eigenstates of the Hamiltonian of QCD, completed with 
the two lowest-order QED interactions of these particles
and photons, $\gamma$, is written as~\cite{BD}
\beq
H \es H_0 + H_1 + H_2 \ .
\eeq
The subscripts denote powers of electric charge.
In the intuitive notation, relevant terms are
\beq
H_0 \es H_\gamma + H_{q \bar q S} + H_p  \ , \\
H_1 \es H_{q \bar q S\, \gamma} + H_{\gamma \, q \bar q S} 
        + H_{\gamma p' \, p} + H_{p \, \gamma p'}   \ , \\
H_2 \es H_{q\bar q S p' \, p} + H_{p \, q\bar q S p'} + \delta H_p \ .
\eeq 
The term $\delta H_p$ denotes the proton electromagnetic 
self-energy counter-term, {\it cf.}~\cite{GellMannGoldberger}. The 
proton eigenstate of $H$ is written as a superposition of just
three components, neglecting terms that eventually do not 
contribute to the string probability at order $e^4$, 
\beq
|{\rm proton}\rangle \es N_{\rm proton} \left( \ |p\rangle + |\gamma p'\rangle + |q\bar q S \, p' \rangle \ \right) \ .
\eeq
It is normalized by the constant $N_{\rm proton}$ so that 
$\langle {\rm proton}|{\rm proton}\rangle = E_p V $, 
where $V$ is the large volume in laboratory frame,
in which the quantum theory is developed. Including proton 
spin, $\langle {\rm proton} \ s|{\rm proton}' \ s'\rangle 
= 2E_p (2\pi)^3 \delta^3(p-p') \delta_{s s'}$. In the formal
scattering theory~\cite{GellMannGoldberger}, the eigenvalue 
equation, 
\beq
H |{\rm proton}\rangle \es E_p |{\rm proton}\rangle \ ,
\eeq
determines the amplitude of string component in the form
\beq
\label{stringA} 
&&
\langle q \bar qS \, p' |{\rm proton} \rangle  
\rs
{ N_{\rm proton} \over E_p - E_{p'} - E_{q \bar q} + i\epsilon  } 
\nt
\sum_\gamma
{
\langle q \bar qS | H_{q \bar q S \, \gamma} | \gamma \rangle 
\
\langle 
\gamma p' | H_{ \gamma p' \, p} |p \rangle
 \over E_p  - E_{p'} - E_\gamma^- + i\epsilon} 
\np
{ N_{\rm proton} \
\langle  q \bar qS \, p'  | H_{q \bar q S \, p' \, p} |p \rangle 
\over E_p - E_{p'} - E_{q \bar q} + i\epsilon  } 
\ . 
\eeq
The matrix elements on the right-hand side are
defined by the assumption that once the quark
pair is created in a pointlike event according to 
QED, the quark and anti-quark move away from 
each other and stretch the string $S$. Thus,
the matrix element $\langle q \bar qS | 
H_{q \bar q S \, \gamma} | \gamma \rangle$
is equal to the QED amplitude for the point-like 
event of creation of the pair, $\langle q \bar qS | 
H_{q \bar q S \, \gamma} | \gamma \rangle
= \langle q \bar q | H_{q \bar q \, \gamma} 
| \gamma \rangle = e_q \, \bar u_q \gamma_\mu 
v_{\bar q} \  \epsilon^\mu_\gamma \ 16 \pi^3 \delta^3
(p_q + p_{\bar q} - p_\gamma) \ \delta_{c_q c_{\bar q}}$.
Similarly, the Coulomb matrix element $ \langle  
q \bar qS \, p'  | H_{q \bar q S \, p' \, p} |p \rangle $ 
is equal to the QED amplitude $ \langle  q \bar q \, p'  | 
H_{q \bar q \, p' \, p} |p \rangle $ for the transition 
$p \to q \bar q \ p'$.  Assuming that the string is 
stretched independently of the spins, flavors and colors 
of quarks, we have
\beq
\label{stringA1}
&&
\langle q \bar qS \, p' |{\rm proton} \rangle  
\rs
{ N_{\rm proton} \over E_p - E_{p'} - E_{q \bar q} + i\epsilon  } 
\nt
\sum_\gamma
{ \langle q \bar q | H_{q \bar q \, \gamma} | \gamma \rangle 
\
  \langle \gamma p' | H_{ \gamma p' \, p} |p \rangle
\over 
E_p  - E_{p'} - E_\gamma^- + i\epsilon} 
\np
{ N_{\rm proton} \
\langle  q \bar q \, p'  | H_{q \bar q \, p' \, p} |p \rangle 
\over E_p - E_{p'} - E_{q \bar q} + i\epsilon  }  \ .
\eeq

\subsection{ String length and orientation }
\label{kr}

The string is stretched by the quarks along their relative 
momentum $\vec k$ in the SRF. At creation, the quark 
has momentum $\vec k$ and the anti-quark $-\vec k$. 
Their invariant mass is $ \cM_{q \bar q} = 2 \sqrt{ m_q^2 
+ \vec k^{\,2} }$. The outward quark motion is slowed 
down by the buildup of a string. In the holography~\cite{zeta1,zeta2} 
motivated by the AdS/QCD duality idea, the buildup is described by 
the decrease of $\cM_{q \bar q}^2$ and increase of the effective 
potential $U_{\rm eff}(r) = \kappa^4 r^2/4$,  where $r$ is 
a three-dimensional distance between the quark and anti-quark 
in the SRF. The quadratic FF holography potential corresponds 
to the linear quark-anti-quark potential that describes the 
gluon strings in the IF~\cite{oscillator}. In the WKB approximation, 
in the FF and IF of Hamiltonian dynamics equally, the quarks 
can reach the distance $r_{\rm max}$ for which the pair potential 
energy equals the initial energy of the quarks' relative motion. This 
implies $r_{\rm max} = 4 |\vec k|/\kappa^2$. The expectation 
value of a string length in the quantum oscillator is smaller 
$\sqrt{2}$ times. Thus, the length and orientation of the 
string $S$ stretched between the quark and anti-quark that 
are created with invariant mass $\cM_{q \bar q}$ from a 
photon, is estimated to be
\beq
\label{rk}
\vec r \es { \sqrt{8} \over \kappa^2 } \ \vec k \ . 
\eeq
With the proton beam along the $x$-axis and the string 
moving very fast along the beam, the string shape of
Fig.~\ref{fig:shape} is built around the vector
\beq
\label{stringrT}
\vec r_T \es { \sqrt{8} \over \kappa^2 } \ (k_y, k_z) \ ,
\eeq
whose azimuthal angle around the beam, measured
from the $y$-axis, is $\varphi = \arctan{(k_z/k_y)}$.
The pair invariant mass and the string length 
in the TP are related through
\beq
\label{pairmass}
\cM_{q \bar q}^2 = 4 ( m_q^2 + \kappa^4 \vec r^{\,2}/8 ) \ .
\eeq 
The string width $w$ is left as a parameter. Strings 
spanned quickly by quarks in photons may be thinner 
than in mesons, because the relative momentum of 
quarks created from a photon in a point-like event is 
not limited, while the relative momentum of quarks in 
typical mesons corresponds to the scale of $\Lambda_{QCD}$. 
For as long as the relativistic mass of quarks, $\gamma m_q $, 
is large, the string is stretched with the speed of light 
irrespective of any dynamical widening that may occur 
later. The strings in photons may be thinner than the 
strings that connect quarks to di-quarks in 
nucleons~\cite{Bjorken:2013boa}, since di-quarks 
are extended objects.

Once the string extension is parameterized by the vector
$\vec r$ that is proportional to $\vec k$ with a fixed
coefficient, the integration over all lengths and directions 
of the strings in $pp$ collisions is equivalent to the integration 
over the relative momenta of quarks in the SRF. In the case of 
fast motion of a string along the beam, in which $\vec r = 
(r_x, \vec r_T)$,  we have
\beq
\label{stringintegral1}
\int dr_x \ \int d^2 r_T  
\es 
{ 8 \sqrt{8} \over \kappa^6 } \int   dk_x \ \int   d^2 k_T  \ .
\eeq

\subsection{ Limitation of string length }

The holographic estimate of string length does not 
include any limitation despite the fact  that gluon strings must 
break before they can reach lengths much greater 
than the hadronic size. Therefore, the integrals in
Eq.~(\ref{stringintegral1}) ought to be limited. 
Although various hypotheses concerning the limit on 
string length can be developed, we find it most 
instructive to set that limit as a model parameter.  
Its magnitude may be on the order of one to ten fm. 
The simplest way for introducing the limitation is to 
cut the integral off. In Eq.~(\ref{stringintegral1}), 
we adopt
\beq
\label{stringintegralLimit}
\int d^3 r & \to & 
\int d^3 r \ \theta( L - |\vec r \, | ) \ ,
\eeq
where $L$ is the length that strings in photons cannot 
exceed. Figure~\ref{fig:example} is obtained using $L 
= 10$ fm, which for the holographic $\kappa \sim 0.5$ 
GeV implies formation of strings of mass up to about 
4.5 GeV. 

The string is stretched off-shell. Therefore, it may be 
active over a period limited by the inverse of its off-shellness. 
Thus, the potentially longer a string the less time for quarks 
to stretch it. In terms of the invariant mass, $\cM$, of a pair 
of light quarks, the time available for string stretching in 
the SRF is $\sim 1/\cM$. Heavy quarks have much less time 
to stretch a string than the light ones have.

In the holographic harmonic oscillator potential that describes 
the string, the quarks slow down. Assuming that the holographic
oscillator frequency is $\omega_{\rm holl} = \kappa^2/(2 m_q)$,
a classical distance at time $t$ is $r(t) = r_{\rm max} \sin 
\omega_{\rm holl} t$. If the available time is $t \sim 1 / \cM$,
the distance between the quark and anti-quark is 
\beq
r(t) \es r_{\rm max} \ \sin \left[ {\kappa^2\over 2 m_q} {1 \over \cM } \right] \ .
\eeq
It is clear that for light quarks there is enough time to reach 
$r_{\rm max}$. For heavy quarks, the FF holography awaits 
justification, but if one assumes the harmonic potential to be 
valid~\cite{HO}, there may not be enough time for charmed 
quarks to reach $r_{\rm max}$. Bottom quarks appear for even 
shorter times. A simplification adopted in estimates made below 
is to ignore the off-shell-time limitation. 

\subsection{ Elements of string probability density }
\label{elements}

\begin{figure}[ht!]
          \includegraphics[width=0.45\textwidth]{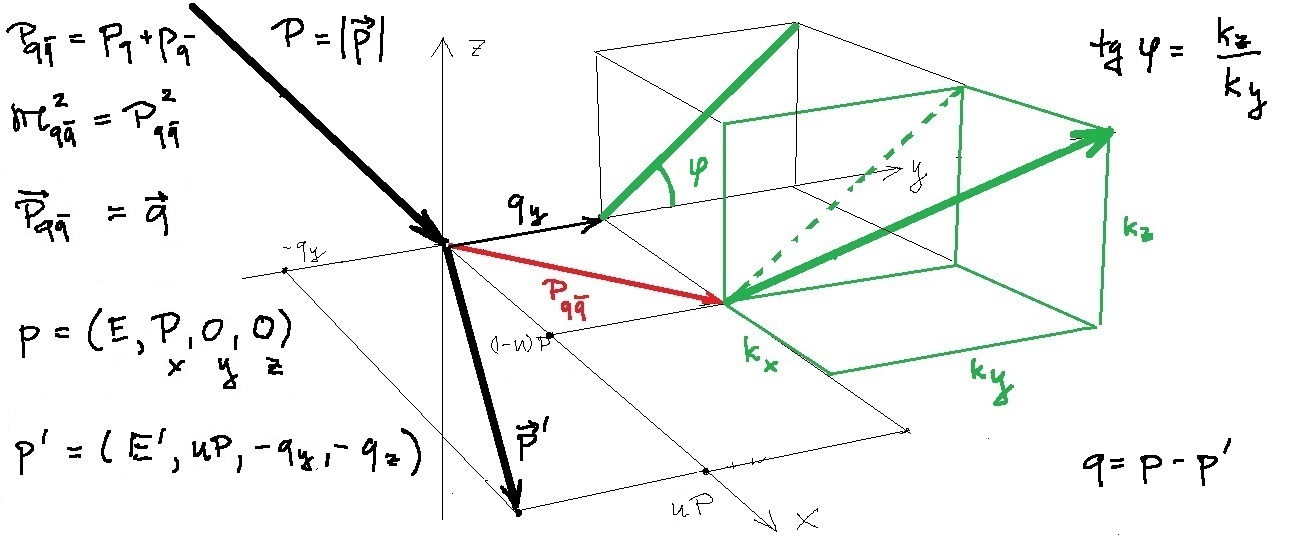}
          \caption{ The proton produces a pair in the Fool's ISR (FISR) frame at 
          LHC. In this example, the proton plane matches the $xy$ plane. The 
          photon emitted by the proton carries the transverse momentum $q_T 
          = (q_y, q_z)$. The figure is simplified by setting $q_z =0$. Generally, 
          $\vec q = (q_x, q_y,q_z) = [(1-u)P,q_y,q_z]$, and the outgoing proton 
          carries $\vec p\, ' = (p'_x,p'_y,p'_z) = (u P, - q_y, -q_z)$. The relative 
          momentum of a quark with respect to an anti-quark is $\vec k$, in the 
          rest frame of the pair. The string between quarks has azimuthal angle 
          $\varphi = \arctan(k_z/k_y)$.  For further details, see Sec.~\ref{scatteringofprotons}.
          \label{FISRstringView}}
          \end{figure}

The matrix element of Eq.~(\ref{stringA1}) is a product of the
proton state of norm squared $2E_p V$ with a quark-anti-quark-proton
state of norm squared $2^3 E_q E_{\bar q} E_{p'}	 V^3$.
In a small volume $d^3p$ there are $d^3 n (2\pi)^3/V$
states. So, the probability of finding a quark pair with a string, $q\bar q S$, 
and a final proton, $p'$, in the initial proton, $p$ with momentum 
quantum numbers $\vec n_p$, in the small volumes of quantum 
numbers around the momentum quantum-number vectors $\vec n_q$, 
$\vec n_{\bar q}$ and $\vec n_{p'}$, is 
\beq
d P
\es
{| \langle q \bar qS \, p' | {\rm proton} \rangle  |^2  \over 2E_p V}
\ {d^3 p_q  \ d^3 p_{\bar q}  \ d^3 p'  
\over 2E_q  \, E_{\bar q} \, E_{p'} [2(2\pi)]^3 } \ .
\eeq
According to Eq.~(\ref{stringA1}), one can write 
\beq
&& 
\langle q \bar qS \, p' | {\rm proton} \rangle  
\rs
N_{\rm proton} \ 
{ (2\pi)^3 \delta^3(p_q + p_{\bar q} - q)  \over  E_p - E_{p'} - E_{q \bar q} + i\epsilon  } 
\nt
\ {-  e_q j^{\mu}_{ q \bar q \, \gamma } \
       g_{\mu \nu} \ 
       e_p j^{\nu}_{\gamma p' \, p} \ \delta_{c_q c_{\bar q}} 
\over  2E_\gamma (E_p  - E_{p'} - E_\gamma + i\epsilon) } \ , 
\eeq
where $q = p - p'$ and
\beq
j^{\mu}_{ q \bar q \, \gamma }
\es
\bar u_{p_q}  \gamma^\mu v_{p_{\bar q}} \ , \\
j^{\nu}_{\gamma p' \, p}  
\es  
\bar u_{p'} 
\left[ \gamma^\nu  F_1(q^2) +  i \sigma^{\nu  \alpha } q_\alpha  { F_2(q^2) \over 2m_p} \right] 
u_p \ .
\eeq
Hence, in the limit of very large initial 
proton momentum $p_x=P$, $p'_x = uP$,
\beq
d P
\es
N_{\rm proton}^2 
\ 2 P(1-u)^2
\ (2 \pi)^3 \delta^3(p_q + p_{\bar q} - q) \ e_p^2 e_q^2
\nt
\ \Bigg| { j^{\mu}\, _{ q \bar q \, \gamma } \ j_{\mu \ \gamma p' \, p}\over (q^2 - \cM_{q \bar q}^2) \ q^2 } \Bigg|^2 \delta_{c_q c_{\bar q}}
\
{d^3 p_q \ d^3 p_{\bar q} \ d^3 p' \over E_q \, E_{\bar q} \, E_{p'} [2(2\pi)]^3 } \ .
\eeq
The kinematics is described in Sec.~\ref{scatteringofprotons} 
and illustrated in Fig.\ref{FISRstringView}.
Integration over the pair total momentum renders 
\beq
d P
\es 
N_{\rm proton}^2
2(1-u)
\ e_p^2 e_q^2
\Bigg| { j^{\mu}\, _{ q \bar q \, \gamma } \ j_{\mu \ \gamma p' \, p}
\over (q^2 - \cM_{q \bar q}^2) \ q^2 } \Bigg|^2 \delta_{c_q c_{\bar q}} 
\nt
\ { d^3 k   \over  (2\pi)^3 \cM_{q \bar q} }
\ {d^3 p' \over 2E_{p'} (2\pi)^3 } \ .
\eeq
A sum over colors yields a factor 3 instead of $\delta_{c_q c_{\bar q}}$.
After averaging over initial proton spins and summing
over final proton spins and quark spins and flavors, with $e_p=1$,
\beq
d P
\es
N_{\rm proton}^2
3  \sum_{\rm flavor} (e_q/e_p)^2\ (1-u)
\nt
{  T(x ,k^\perp, p', p) 
\over
(q^2 - \cM_{q \bar q}^2)^2 \ q^4 } 
\ { d^3 k \over (2\pi)^3 \cM_{q\bar q} }
\ {d^3 p' \over 2E_{p'} (2\pi)^3 } \ ,
\eeq
where
\beq
T(x ,k^\perp, p', p) 
\es
\sum_{s_p} \sum_{s_q s_{\bar q} s_{p'}}  
\big| j^{\mu}\, _{ q \bar q \, \gamma } \ j_{\mu \ \gamma p' \, p} \big|^2  \ .
\eeq
Evaluation yields
\beq
\label{traceformula1}
&& T(x ,k^\perp, p' , p) 
\rs  
16 (F_1+F_2)^2  \ \cT_f
\nm
16 \ F_2 \left[ F_1 + \h \left( 1 + {q^2 \over 4m_p^2} \right) F_2 \right] \ \cT_s \ , 
\eeq
with
\beq
\label{traceformula2}
\cT_f 
\es
2(t_1 \, t_4 + t_2 \, t_3)
- m_p^2 \cM^2_{q \bar q} - q^2 m_q^2 \ , \\
\label{traceformula3}
\cT_s 
\es
2(t_1 \, t_4 + t_2 \, t_3)
+ 
2(t_1 \, t_2 + t_3 \, t_4)
\np
( q^2 - 4m_p^2) \cM_{q \bar q}^2/2 \ , \\
\label{traceformula4}
t_1 \es pp_q 
\ , \ 
t_2 \rs pp_{\bar q} 
\ , \   
t_3 \rs p'p_q 
\ , \  
t_4 \rs p'p_{\bar q} \ .
\eeq
The proton form factors $F_1$ and $F_2$ are 
parameterized as in Ref.~\cite{F1F2}. 

Using notation $[p'] = d^3 p' /[2E_{p'} (2\pi)^3]$
and Eqs.~(\ref{rk}) and (\ref{stringintegralLimit}),
the string probability density in the space of vectors 
$\vec r$ is obtained in the form
\beq
\label{Prpp}
&&
P(\vec r, p',p)
\rs
{d P \over \, d^3r} 
\rs
[p'] \
N_{\rm proton}^2 
\ { 3  \ \kappa^6 \over 8 \sqrt{8}  \ (2\pi)^3 } \
\nt
\sum_{\rm flavor} {e^2_q \over e_p^2}  \ {1-u\over   \cM_{q\bar q} } \
{ T(x ,k^\perp, p' , p)  
\over
(q^2 - \cM_{q \bar q}^2)^2 \ q^4 } 
\
\theta(L - r )
\ .
\eeq 
Its magnitude includes the volume of the final 
proton detector momentum bin, $[p']$.
Formally, the same formulas are obtained 
using the FF and IF of dynamics, except for
the difference in direction in the off-shell
continuation. The same result holds for
electrons, when one sets $F_1=1$, $F_2=0$
and replaces the proton mass $m_p$ by the
electron mass $m_e$. 

The probability $P(\Phi)$ in Fig.~\ref{fig:PPhi} is 
obtained by introducing spherical co-ordinates,
$ \vec r  = ( r \sin \beta , r \cos\beta  \cos \varphi , 
r \cos\beta  \sin \varphi) $ with $\beta = \pi/2 - \theta$
and $\theta$ measured from $x$-axis, which is the 
beam direction, so that $d^3 r = r^2 dr \ \sin \theta 
\ d\theta \ d\phi$. One integrates $\int_0^L r^2 dr$ 
and $\int_0^\pi \sin \theta \, d\theta$ and sets 
$\phi = \Phi$. 

\section{ String collisions in $pp$ scattering  }
\label{scatteringofprotons}

We consider peripheral $pp$ scattering that proceeds
through photons. In the LHC laboratory frame of reference, 
the $x$-axis is set along the proton beams, $z$-axis is 
vertical and $y$ axis is  along $\hat z \times \hat x$. 

\subsection{ Proton momenta }

Initial protons have four-momenta 
\beq
\label{initialIF1}
p_1 \es ( E , + P , 0 , 0 ) \ , \\ 
\label{initialIF2}
p_2 \es ( E , - P , 0 , 0 ) \ ,
\eeq
with $E = \sqrt{m_p^2 + P^2}$. Final protons 
four-momenta,
\beq
p'_1 \es ( E'_1 , \  p'_{1x} , \ p'_{1y} , \ p'_{1z} )  \rs p_1 - q_1  \ , \\ 
p'_2 \es ( E'_2 , \  p'_{2x} , \ p'_{2y} , \ p'_{2z} )  \rs p_2 - q_2  \ ,
\eeq
include energies 
\beq
E'_1 \es \sqrt{   m_p^2  + (P - q_{1x})^2 + q_{1y}^2 + q_{1z}^2 } \ , \\
E'_2 \es \sqrt{   m_p^2  + (- P - q_{2x})^2 + q_{2y}^2 + q_{2z}^2 } \ ,
\eeq 
and the azimuthal angles  $\Phi_1$ and $\Phi_2$ of the proton planes
\beq
\vec p\,'_{T1} \es ( - q_{1y} , \ - q_{1z} ) \rs p'_{T1} ( \cos \Phi_1, \sin \Phi_1) \ , \\
\vec p\,'_{T2} \es ( - q_{2y} , \ - q_{2z} ) \rs p'_{T2} ( \cos \Phi_2, \sin \Phi_2) \ .
\eeq
The angle $\Phi$ in Fig.~\ref{fig:example} equals 
$\Phi_1 - \Phi_2$.
The protons are assumed to lose a sizable fraction 
of their momentum along the beam, so that the 
strings have a lot of energy to produce $X$. So, 
$q_{1x}$ is on the order of $P$ and $q_{2x}$ is 
on the order of $-P$. Denoting both in-coming 
protons momenta equally by $p$ and both photon 
momenta equally by $q$, we have ${p'}_x = 
\pm P  -  q_x =  \pm u P$ and in the limit
$P \to \infty$,
\beq
q^2 \es - \ { (1-u)^2 m_p^2 + q_y^2 + q_z^2 \over u } + O(1/P^2)  \ .
\eeq

\subsection{ Quark momenta }

In terms of the SRF quark relative momentum $\vec k$, 
the four-momentum of a quark from proton 1 reads
\beq
p_q^0    
\es 
\h \sqrt{ \cM^2 + \vec q\,^2 } + {\vec q \cdot \vec k  \over  \cM } \ , \\
\vec p_q 
\es 
\h \, \vec q 
+
{ \sqrt{ \cM^2 + \vec q\,^2 } \over \cM } \
{\vec q \cdot \vec k  \over \vec q\,^2 }  \ \vec q
+ 
\vec k - {\vec q \cdot \vec k  \over  \vec q \,^2 } \ \vec q \ , 
\eeq
where $ \vec q = \left[ (1-u) P, \ q_y , \ q_z \right]$
and $\cM = 2 \sqrt{ m_q^2 + \vec k\,^2}$.
Components of the anti-quark four-momentum $p_{\bar q}$ 
are obtained by changing $\vec k$ to $-\vec k$. The change 
of sign in front of $P$ provides expressions for quarks 
coming from  proton number 2. 

\subsection{ Probability $P(\vec r, p',p)$ }

Given the incoming and outgoing particles' momenta,
elements of probability densities in 
Eqs.~(\ref{traceformula1})-(\ref{traceformula4}), are
\beq
\label{Tf}
\cT_f 
\es
{  R_1 R_4 + R_2 R_3  \over  2 x (1-x) u     (1-u)^2  }
- m_p^2 \cM^2_{q \bar q} - q^2 m_q^2 \ , \\
\cT_s 
\es
{  R_1 R_4 + R_2 R_3  \over  2 x (1-x) u     (1-u)^2  }
+ 
{ u^2 R_1 R_2 + R_3 R_4 \over  2 x (1-x) u^2 (1-u)^2  }
\np
( q^2 - 4m_p^2) \cM_{q \bar q}^2/2 \ ,
\eeq
where, using $k = (k_y,k_z)$ and $q = (q_y,q_z)$,
\beq
R_1 \es  (k \hspace{-.8mm}+\hspace{-.8mm} x  q )^2 \hspace{-.8mm}+\hspace{-.8mm} m_q^2 
\hspace{-.8mm}+\hspace{-.8mm}  [x(1\hspace{-.8mm}-\hspace{-.8mm}u)m_p]^2 \ , \\
R_2 \es  [k \hspace{-.8mm}-\hspace{-.8mm} (1\hspace{-.8mm}-\hspace{-.8mm}x) q ]^2 
\hspace{-.8mm}+\hspace{-.8mm} m_q^2 \hspace{-.8mm}+\hspace{-.8mm}
 [(1\hspace{-.8mm}-\hspace{-.8mm}x)(1\hspace{-.8mm}-\hspace{-.8mm}u)m_p]^2 \ , \\
R_3 \es  (u k \hspace{-.8mm}+\hspace{-.8mm} x  q )^2 \hspace{-.8mm}+\hspace{-.8mm} u^2 m_q^2 
\hspace{-.8mm}+\hspace{-.8mm} [x(1\hspace{-.8mm}-\hspace{-.8mm}u) m_p]^2 \ ,\\
R_4 \es  [u k \hspace{-.8mm}-\hspace{-.8mm}(1\hspace{-.8mm}-\hspace{-.8mm}x) q ]^2 
\hspace{-.8mm}+\hspace{-.8mm} [u m_q]^2 \hspace{-.8mm}+\hspace{-.8mm} [(1\hspace{-.8mm}-\hspace{-.8mm}x)(1\hspace{-.8mm}-\hspace{-.8mm}u)m_p]^2\hspace{-.8mm} .
\eeq
Probability $P(\vec r, p',p)$ of Eq.~(\ref{Prpp})
is obtained using Eq.~(\ref{rk}) for $ \vec k$,
so that the string vector $ \vec r  = ( r \sin \beta , 
r \cos\beta  \cos \varphi , r \cos\beta  \sin \varphi) $,
where $\beta$ and $\varphi$ are the angle the string 
forms with the TP and the azimuthal angle measured 
from $y$-axis, respectively. We have
\beq
k_y  \es   (\kappa^2 / \sqrt{8})  \  r \  \cos \beta \cos \varphi \ , \\
k_z  \es   (\kappa^2 / \sqrt{8})  \  r \  \cos \beta \sin \varphi \ , \\  
x     \es  \h  + (\kappa^2 / \sqrt{8})  \  { r  \  \sin \beta  \over \cM_{q \bar q}} \ ,
\eeq
and $\cM_{q \bar q}^2 = 4 ( m_f^2 + \kappa^4  r^{\,2}/8 ) $. 

\subsection{ Characteristics of string collisionsSRF }

Scattering characteristics shown in Fig.~\ref{fig:example} 
are evaluated using Eqs.~(\ref{NMB}) and (\ref{epsMB})
and
\beq
\label{NppC0-Phi}
{ N(\Phi) \over N(0)} \es { N(p'_1,p'_2)  \over N(p'_{10},p'_{20}) } 
\rs 
{ \tilde N_{\rm coll} (p'_1,p'_2)  \over \tilde N_{\rm coll}(p'_{10},p'_{20}) }
\ , \\
\label{epsppC0-Phi}
{v_2(\Phi) \over v_2(0) }
\es
{ \varepsilon_2(p'_1,p'_2)  \over \varepsilon_2(p'_{10},p'_{20}) } 
\rs
{ \tilde \varepsilon_2(p'_1,p'_2) \over \tilde \varepsilon_2 (p'_{10},p'_{20}) } \ ,
\eeq
where the functions $\tilde N_{\rm coll} (p'_1,p'_2)$ 
and $\tilde \varepsilon_2(p'_1,p'_2)$ do not contain 
any constant factors that cancel out in the evaluated 
ratios. These functions are defined in App.~\ref{tildes},
Eqs.~(\ref{NppC}) and (\ref{epsppC}). Their values
result from integration over the range of impact vectors 
$\vec b$, for which strings are capable of slamming into
each other, on the features of strings overlap area from 
Fig.~\ref{fig:overlap}, and on the probability distribution 
of the string vectors $\vec r_1$ and $\vec r_2$, over 
which one integrates in MB averages. For example, 
strings cannot collide if half of the sum of their lengths 
is smaller than the length of the impact vector, collisions 
of thin strings mostly occur through the rhombus shape 
of Fig.~\ref{fig:intx-overlapcalculation-20171217}, and 
experimental cuts on multiplicity or elliptic flow limit the 
azimuthal angle $\gamma$ between vectors $\vec r_1$ 
and $\vec r_2$. 

Suppose that strings are chains of effective 
gluons~\cite{RGPEPgluons} and the volume of a gluon 
in a chain is $w^3$. The corresponding density of gluons 
is then $\rho = 1/w^3$. Consequently, the cross-section 
for inelastic gluon-gluon scattering is $\sigma \sim w^2$. 
The number of gluons in a string of length $r$ much longer 
than the width $w$, $N_g \sim r/w$, is large, which justifies 
the Glauber-model formula of Eq.~(\ref{dsigma}), which
appears in Eqs.~(\ref{NMB}) and (\ref{epsMB}) and their
computational forms in Eqs.~(\ref{NrrC}) and (\ref{epsrrC}).
The rhombus area in Eq.~(\ref{arhombus}) is independent 
of the impact vector $\vec b$ for any sizable $\gamma$. 
So is the corresponding number of collisions, 
$N_{\rm coll}(\vec b, \vec r_1, \vec r_2)$, which makes
the exponential in the cross-section take the form 
$\exp (- c/| \sin \gamma \, |)$, where for the thin strings
introduced above the constant $c > 1$.

\begin{figure}[ht]
          \includegraphics[width=0.35\textwidth]{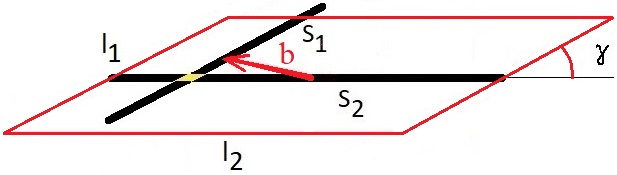}
          \caption{  Illustration of the parallelogram in the transverse plane, 
          over which the end of impact vector $\vec b$ ranges for some fixed 
          values of the strings lengths $l_1$ and $l_2$ and their relative azimuthal 
          angle $\gamma$ in the TP. In almost entire parallelogram, the overlap 
          area and its eccentricity are the same when the ratios $w/l_1$ and $w/l_2$ 
          are negligible. The overlap area is marked in yellow. 
          \label{fig:intx-parallelogramb-20171213} }
          \end{figure}

Integration over the impact vectors is illustrated in 
Fig.~\ref{fig:intx-parallelogramb-20171213}. For 
every choice of string transverse lengths $l_1$, $l_2$ 
and their relative azimuthal angle $\gamma$, the strings 
have a non-zero overlap area when the end of vector 
$\vec b$ lies anywhere within a parallelogram of sides 
$l_1$ and $l_2$.
For all vectors $\vec b$ in the parallelogram, the overlap 
area and its eccentricity are approximately the same, 
variations appearing only near the boundaries of the 
parallelogram and for $\gamma \sim 0$ or $\gamma \sim \pi$. 
Thus, the parallelogram picture applies in the case of string 
width $w$ much smaller than the string lengths $l_1$ and $l_2$.
The area of such a parallelogram is $l_1 l_2 |\sin \gamma \, |$.
Hence, using Eqs.~(\ref{arhombus}) and (\ref{erhombus}),
the thin string approximation for angles $\gamma$ much 
greater than $\gamma_0 = 2w/l_1 + 2w/l_2$  yields
\beq
\label{tildeNrr}
&& \tilde N_{\rm coll}(\vec r_1, \vec r_2)  \\
\es
l_1 \, l_2 \, 
\left[ 1 - e^{- \sigma ( \rho \pi w/4 )^2 \hat \rho_1 \hat \rho_2 w^2 / | \sin \gamma \, |  } \right]  \
\ \hat \rho_1 \hat \rho_2 w^2  \ , \nn
&& 
\tilde \varepsilon_2(\vec r_1, \vec r_2) \\
\es
 l_1 \, l_2 \, | \sin \gamma \, |
\left[ 1 - e^{- \sigma ( \rho \pi w/4 )^2 \hat \rho_1 \hat \rho_2 w^2 / | \sin \gamma \, |  } \right]  
\ | \cos \gamma \ | \ , \nnn
\eeq
where the density factor $\hat \rho$, is defined
in Eq.~(\ref{tilderho}), resulting from Eq.~(\ref{rho}).

Note that after integration over impact vectors the
MB average number of collisions is not inversely 
proportional to  $|\sin \gamma \, |$, as intuition 
for small impact vectors suggests. Similarly, eccentricity 
is not just proportional to $| \cos \gamma \,|$, but 
rather to $| \sin 2 \gamma \, |$. The reason is that 
both these characteristics include the factor 
$|\sin \gamma \,|$ that results from integration 
over impact vectors. Therefore, the MB averaged 
multiplicity and elliptic flow correlations shown in 
Fig.~\ref{fig:example} result from the inverse of 
$\sin \gamma$ in the exponential representing the Glauber-model 
formula. 

When $\gamma \to 0$, the Glauber-model exponential
tends to zero and the rhombus approximation yields 
\beq
\label{NrrCevrhomboid-simple2}
\tilde N_{\rm coll}(\vec r_1, \vec r_2)
\es
\hat \rho_1 \ \hat \rho_2 \ w^2 \ l_1 \, l_2  \ , \\
\label{epsrrCevrhomboid-simple2}
\tilde \varepsilon_2(\vec r_1, \vec r_2)
\es
 l_1 \, l_2 \ | \sin \gamma \ | \, | \cos \gamma \, | \ ,
\eeq
while exact integration for $\gamma = 0$, ignoring 
the exponentials, results in
\beq
\label{Nrrev2g07simple5}
\tilde N_{\rm coll}(\vec r_1, \vec r_2)|_{\gamma=0} 
\es
\hat \rho_1 \ \hat \rho_2 \ w^2 \ l_1 l_2 \ , \\
\label{epsrrev1g07simple5}
\tilde \varepsilon_2(\vec r_1, \vec r_2)|_{\gamma=0} 
\es
4( l_1 l_2 + w^2) \arctan(w/l_1) 
\nm       
2w( l_2 - l_1 + 2 w) \ .
\eeq
In the limit $\gamma \to 0$, the two results for 
$\tilde N_{\rm coll}$ agree well with each other. 
In contrast, the rhombus 
approximation for eccentricity does not extend below 
the minimal angle order $w/l$ at which the rhombus 
shape is altered by interference from the ends of strings. 
The eccentricity formula of Eq.~(\ref{epsrrCevrhomboid-simple2}),
valid for $\gamma \gg \gamma_0 = 2w/l_1 + 2w/l_2$,  does 
not automatically match the exact result for $\gamma = 
0$, with neglected exponentials. Instead, a simple 
formula achieves the required matching,
\beq
\tilde \varepsilon_2(\vec r_1, \vec r_2)
\es
l_1 \, l_2 \ \left( \, | \sin \gamma \, | +  \gamma_0 \right) \  | \cos \gamma | \ . \nnn
\eeq
Hence, for general angles $\gamma$, the thin string 
model provides the following expressions that are used
in our estimates 
\beq
\tilde N_{\rm coll}(\vec r_1, \vec r_2)
\es
w^2 \ (r_1+w) \  (r_2+w) \nt
\left[ 1 - e^{- \sigma ( \rho \pi w/4 )^2 \hat \rho_1 \hat \rho_2 w^2 / | \sin \gamma \, |  } \right]\ , \\
\tilde \varepsilon_2(\vec r_1, \vec r_2)
\es
l_1 \, l_2 \ \left( \, | \sin \gamma \, | +  \gamma_0 \right) \  | \cos \gamma |  \nt
\left[ 1 - e^{- \sigma ( \rho \pi w/4 )^2 \hat \rho_1 \hat \rho_2 w^2 / | \sin \gamma \, |  } \right]\ .
\eeq
We use these formulas to obtain the ratios shown 
in Fig.~\ref{fig:example} from Eqs.~(\ref{NMB}) 
and (\ref{epsMB}), or their equivalents (\ref{NppC0-Phi}) 
and (\ref{epsppC0-Phi}). The six-dimensional integrals 
over string vectors $\vec r_1$ and $\vec r_2$ are 
carried out numerically.

\subsection{ Beyond minimal bias }
\label{BMB}

A condition of relatively large multiplicity selects events
with small values of the angle $\gamma$ in the minimal 
bias expectation values only through the Glauber 
cross-section exponential, and not directly through 
the overlap shape. This is so because the $\sin \gamma$ 
that results from integration over impact vectors $\vec b$ 
cancels the effect of rhombus overlap area growing like 
$1/ \sin \gamma$ for small angles $\gamma$. In  bins 
of data with large multiplicity, for which the Glauber 
exponential ought to be small, there also ought to be an 
increase in elliptic flow.

Indeed, this is what happens according to Eqs.~(\ref{NMB}) 
and (\ref{epsMB}) when one limits the averaging to
events with large multiplicities. The resulting effects 
can be estimated by requiring that the number of binary 
collisions of Eq.~(\ref{tildeNrr}) in the events over which 
one averages is not smaller than a specified fraction $f$ 
of the maximal observed multiplicity in the peripheral events. 
According to Eq.~(\ref{tildeNrr}), this condition reads
\beq
\tilde N_{\rm coll}(\vec r_1, \vec r_2) 
\ge
f \  w^2 (L + w)^2  \ .
\eeq
Computation shows that for $f = 1/2$ and $f=0.75$ 
the multiplicity ratio $N(\pi/2)/N(0)$ decreases from 
its MB value of about 0.975 in Fig.~\ref{fig:example},  
to about 0.93 and 0.92, respectively. This is nearly a
three times bigger oscillation. Correspondingly, the 
elliptic flow ratio $v_2(\Phi)/v_2(0)$ drops from its 
MB value of about 0.965 in Fig.~\ref{fig:example} to  
about 0.90 and 0.875, which is even bigger than a 
threefold increase in the amplitude. 

\subsection{ Sensitivity to string features }

Our assumed gluon string width $w \sim 0.1$ fm is extremely small. 
In our estimates, when this is decreased or increased by a
factor of three, the multiplicity ratio at its minimum at 
$\Phi = \pi/2$  does not visibly change. The eccentricity 
ratio at its minimum decreases by about a percent when 
$w$ increases in that range. 

The upper limit we imposed on the string length, 
$L=10$ fm, also appears extreme. Reduction of $L$
by half causes an increase of the multiplicity ratio at 
its minimum to 99\% and an increase of the 
eccentricity ratio at its minimum to about 98\%.
Thus both azimuthal correlation effects become 
approximately halved. 

Theoretically, when the impact parameter $b$ in a $pp$ 
ultra-peripheral collision exceeds some value $b_{\rm min}$
larger than the strong-interaction proton diameter, one
may exclude from the integration range over $\vec b$ in 
Fig.~\ref{fig:intx-parallelogramb-20171213} a circle of 
radius $b_{\rm min}$.  For example, when $b_{\rm min}$
is set to 4 fm, and the product of string lengths $l_1 l_2$
is replaced by $l_1 l_2 - \pi b_{\rm min}^2 > 0$, the multiplicity 
and eccentricity ratios at their minima drop to about 0.93,
increasing the string azimuthal effect. This can be interpreted 
as a consequence of the feature that only long strings can
collide for large $b$ and for long strings the azimuthal 
correlation may be more pronounced than for short ones.

A line of study emerges regarding variation of the multiplicity 
and elliptic flow ratios with the projectile momentum transfers,  
such as the variation of Fig.~\ref{fig:example} with the photon 
$q_1^2$ and $q_2^2$ in $pp \to ppX$ collisions. For example, 
when one of the protons referenced in Fig.~\ref{fig:example} 
loses half instead of a quarter of its initial momentum along the 
beam, which implies that its momentum transfer squared changes 
from $-1.4$ to  $-2.4$~GeV$^2$, the ratios of Fig.~\ref{fig:example} 
increase at their minima by about one percent. 

\subsection{ Comparison with QED lepton-pair production }
\label{QEDpairs}

In our study, the initial stage of a string-string collision   
resembles photon-photon production of quark pairs. 
Virtual photons emitted from the beam particles dominate 
the forward scattering in which one can study string 
signatures. Key to the formation of these signatures are 
the initial-stage correlations among the two scattering 
planes of beam particles and two quark-pair planes. We 
explain the relevance here of the analogous QED results 
for the production of two lepton pairs with proton beams. 
Each virtual photon coming from its respective proton 
turns into a pair of leptons and the two pairs exchange 
another virtual photon.

We can infer some preliminary idea of the QED results from 
Ref.~\cite{RefA}. In the on-shell reaction, $\gamma 
\gamma \to e^+ e^- + e^+ e^-$, the cross section is dominated 
by the collinear mass divergences arising from the electron and 
photon propagator poles. This in turn implies that each pair is 
close to threshold with little average azimuthal correlation between 
the pair planes, for unpolarized real photons. This remains true 
for virtual photons whose average includes the longitudinal mode 
and whose spacelike mass is small compared to the pair energies. 
Taking into account the projectile current, on the other hand, we 
find correlations between the projectile scattering plane and its 
corresponding lepton-pair plane. What is of primary interest is 
the correlation between the two lepton-pair planes, and how 
this correlation depends on the angle between the projectile 
planes. 

Comparison of the string-string collision correlations with QED 
lepton pair production hinges on the degree to which a photon 
exchange between lepton pairs and the strong-interaction 
between quark-pairs differ in their influence on the final-particle 
distributions. As a benchmark for comparisons, the QED results by 
themselves are of interest in all of the three cases of proton beams 
induced, electron beams induced or electron beam and proton beam 
induced double lepton pair production. Although the QED discussion 
is beyond the scope of the present paper, we wish to mention that 
the benchmark estimates can be based on a combination of analytical 
and numerical calculations as in Ref.~\cite{RefA} or on the 
event-generator approach as in Ref.~\cite{RefB}. The central need 
is to identify the data cuts for simulations that allow for comparison 
with the string model predictions. 

\section{ Conclusion }

As we have shown in this paper, the physics of ridge production seen in high multiplicity hadronic events in proton-proton collisions at the LHC  has important consequences for high energy collisions mediated by photons. This includes ultra-peripheral $p p $ collisions at the LHC, photon-photon collisions at a high energy electron-positron collider, and electron-proton collisions at an electron-ion collider. In each case the virtual photon creates a quark-antiquark pair connected by a gluonic string, {\it i.e.}, a gluon flux tube. 

The gluonic string represents the QCD dynamics of the force which confines the triplet and anti-triplet 
color of the quark and antiquark.   For example, the confining harmonic oscillator  potential in the 
light-front holographic model~\cite{zeta1,zeta2} can be identified with the dynamics of a gluon string. 

In the case of an electron at an electron-ion collider,  the frame-independent wavefunction of its 
$|e^-  q \bar q \rangle  $ Fock state,  defined at fixed light-front time $x^+ = t + z/c$,  is off-shell in $P^- = P^0 + P^z$ and thus off-shell in the invariant $q \bar q$  mass. The quark and antiquark in the Fock state are confined  via the exchange of gluons, the same string-like interactions responsible for color confinement and Pomeron exchange.  The virtual state of the lepton becomes on-shell in the electron-proton collision.  The high energy collisions of the two flux tubes will produce maximal hadronic multiplicity when the flux tubes are maximally aligned, {\it i.e.}, when the area of overlap in the transverse plane is maximal as illustrated in Fig.~\ref{fig:intx-overlapcalculation-20171217}.  Moreover, as we have shown,  the azimuthal distribution of the resulting hadronic ridges  will be correlated with the scattering plane of the scattered lepton as 
illustrated in Fig.~\ref{fig:example}. 

The ratios of multiplicities and elliptic flows shown in 
Fig.~\ref{fig:example} are independent of absolute
probabilities or cross sections for the events they 
concern. The absolute quantities cannot be estimated 
without using advanced models~\cite{Baltz}. The ratio 
of cross sections for single tagged versus untagged 
$p p \to X$ hadronic events at the LHC and RHIC are
needed. One needs to estimate the event rate for 
high multiplicity events in $\gamma p $ collisions and the
analogous quantity for the process $pp \to ppX$.
High-multiplicity cuts deplete the number of available
events by a factor $10^{-6}$ or smaller \cite{CMScollectivity}
in central collisions. In peripheral ones, additional factors
of powers of $\alpha \sim 1/137$ significantly reduce the 
probability to see large number of products in the final
state $X$. 

However, it is possible to consider  replacement
of a photon by a pomeron. A string due to a photon 
from a lepton or a proton may collide with a string due 
to the pomeron from another proton. Instead of the factor
$\alpha^4$, one obtains the much larger value $\alpha^2$. In 
such a setup, in analogy to deep inelastic $ep$ scattering 
illustrated in Fig.~\ref{fig:ep}, one could only seek 
alignment of elliptic flow with the electron scattering plane. 
Single-hadron correlation with a projectile plane, in that
case a lepton, has already been studied~\cite{ZEUSazimuthal}.
However, azimuthal asymmetry of elliptic flow in the final
state has to our best knowledge not been measured.
If two pomerons replaced two photons, the small factors 
due to $\alpha$ would be eliminated. However, note that 
LHC is already considered as a photon-photon collider and 
software for simulating exclusive production is being 
built~\cite{Harland-Lang:2017cax}. As far as we know, 
extension to string collisions due to the photons has not 
been considered yet.

As a final warning, it should be kept in mind that observable 
many-body effects due to collisions of gluonic strings are not 
guaranteed to be describable by many-body techniques used 
for nucleons in heavy ion physics~\cite{eccentricity1,eccentricity2,eccentricity3}.
Quarks and gluons are confined objects whose interactions
at the distances that characterize their binding mechanism
and even the larger distances at which strings may form, 
are much less understood than the interactions of nucleons 
are in the context of formation of nuclei and physics of nuclear 
reactions. Discussion of implications of the string picture 
for scattering processes that involve ions, including dependence 
on atomic numbers A and Z from one to the largest available 
values, would require extension of the theory.

In summary, our estimates for collisions of gluon strings 
suggest that building required theory and computational 
tools for absolute estimates should follow experimental 
verification if the azimuthal variations of multiplicity and 
elliptic flow do manifest themselves in measurements of 
ratios exemplified in Fig.~\ref{fig:example}. Thus, instead 
of predicting the absolute size of multi-particle string-collision
effects, our estimates pose a question of to what extent the 
correlations of the type illustrated in Fig.~\ref{fig:example} 
do actually occur in proton-proton, lepton-proton or even 
lepton-lepton scattering. Experimental assessment of their 
magnitude would motivate directions for developing theory
and studying gluon strings using LHC and other machines.

\vskip.2in

{\bf Acknowledgment }
\vskip.1in

The authors thank James Bjorken, Dmitri Kharzeev, 
Edward Shuryak, Ramona Vogt and Ismail Zahed 
for discussions. This work is also supported by the 
Department of Energy contract DE--AC02--76SF00515. 

\begin{appendix}
\section{ Eqs.~(\ref{NMB}) and (\ref{epsMB}) }
\label{tildes}

Equations~(\ref{NMB}) and (\ref{epsMB}) for average 
string-collision quantities and Eqs.~(\ref{NppC0-Phi})
and (\ref{epsppC0-Phi}) for the ratios shown in 
Fig.~\ref{fig:example}, involve integration over 
string vectors $\vec r_1$ and $\vec r_2$  as arguments of 
the string probability densities $P(\vec r_1, p'_1,p_1)$ and 
$P(\vec r_2, p'_2,p_2)$. Since the densities are independent 
of the impact vector $\vec b$, the order of integration in 
Eqs.~(\ref{NMB}) and (\ref{epsMB}) can be changed and
multiplicity $N$ and eccentricity $\varepsilon_2$ are
\beq
\label{Npp}
&& N(p'_1,p'_2) 
\rs 
{C_N \over \sigma_{\rm p } }
\int d^3r_1 \ P(\vec r_1, p'_1,p_1)
\nt
\int d^3r_2 \ P(\vec r_2, p'_2,p_2)
\ 
N_{\rm coll} (\vec r_1, \vec r_2) \ , \\
\label{epspp}
&& \varepsilon_2(p'_1,p'_2) 
\rs 
{1 \over \sigma_{\rm p } }
\int d^3r_1 \ P(\vec r_1, p'_1,p_1)
\nt
\int d^3r_2 \ P(\vec r_2, p'_2,p_2)
\ \varepsilon_2(\vec r_1, \vec r_2) \ , 
\eeq
where
\beq
\label{Nrr}
&& N_{\rm coll} (\vec r_1, \vec r_2)
\rs
\int d^2b 
\left[ 1 - e^{-N_{\rm coll}(\vec b, \vec r_1, \vec r_2)} \right]  
\nt 
N_{\rm coll}(\vec b, \vec r_1, \vec r_2) \ , \\
\label{epsrr}
&& \varepsilon_2(\vec r_1, \vec r_2)
\rs
\int d^2b 
\left[ 1 - e^{-N_{\rm coll}(\vec b, \vec r_1, \vec r_2)} \right]  
\nt \varepsilon_2(\vec b, \vec r_1, \vec r_2) \ .
\eeq
In the ratios shown in Fig.~\ref{fig:example} all 
constant factors cancel out and they can be 
removed from calculation. With the use of definitions
\beq
\label{NppC0}
N(p'_1,p'_2) 
\es 
C_N  \ C_1  \ C_2 \ \tilde N_{\rm coll} (p'_1,p'_2) \ , \\
\label{epsppC0}
\varepsilon_2(p'_1,p'_2) 
\es 
C_1 \ \tilde \varepsilon_2(p'_1,p'_2) \ ,
\eeq
where $C_2 = \sigma ( \rho \pi w/4 )^2$ and 
\beq
C_1 \es
{1 \over \sigma_p} 
2 [p'_1] \
N_{\rm proton}^2 
\ { 3 e_p^4 \ \kappa^6 \over 8 \sqrt{8}  \ (2\pi)^3 } \
2 [p'_2] 
\nt
N_{\rm proton}^2 
\ { 3  \ \kappa^6 \over 8 \sqrt{8}  \ (2\pi)^3 } \ .
\eeq
The constant $C_N$ is dimensionless, the dimension of $C_1$
is mass to the power eighteen, and  the dimension of $C_2$ is 
mass squared. Dimensional considerations ought to 
include the fact that all the observables refer to specific 
states of final protons and as such are actually densities 
in the space of final proton momenta, where the measure 
is $[p'_1 p'_2]$ of dimension mass to fourth power. 
The infinitesimal momentum volumes are to be replaced 
with the experimental ranges of detection of the two 
outgoing protons. Other factors are: $\sigma$, 
the parton-parton cross-section; $\rho$, the parton 
density in a string volume; $w$, the string width; 
$N_{\rm proton}$, the dimensionless proton-state 
normalization constant, $e_p=1$, the proton 
charge; and $\kappa$, the holography effective-potential 
parameter $\sim 0.5$ GeV. 

With constants factored out, evaluation of ratios 
in Fig.~\ref{fig:example} can be carried out 
replacing $N_{\rm coll} (p'_1,p'_2)$ and 
$\varepsilon_2(p'_1,p'_2)$ by 
\beq
\label{NppC}
&& \tilde N_{\rm coll} (p'_1,p'_2) 
\rs 
\int d^3r_1 \ \tilde P(\vec r_1, p'_1,p_1)
\nt 
\int d^3r_2 \ \tilde P(\vec r_2, p'_2,p_2)
\ \tilde N_{\rm coll} (\vec r_1, \vec r_2) \ , \\
\label{epsppC}
&& \tilde \varepsilon_2(p'_1,p'_2) 
\rs 
\int d^3r_1 \ \tilde P(\vec r_1, p'_1,p_1)
\nt
\int d^3r_2 \ \tilde P(\vec r_2, p'_2,p_2)
\ \tilde \varepsilon_2(\vec r_1, \vec r_2) \ , 
\eeq
where the probability densities without constants are  
\beq
&& \tilde P(\vec r, p',p)
\rs \theta(L - r ) \nt
\sum_f {e_f^2 \over e_p^2} \ {1-u \over \cM_{q \bar q} } \
{ T(x ,k^\perp, p' , p)  
\over
(q^2 - \cM_{q \bar q}^2)^2 \ q^4 } \ , 
\eeq
and
\beq
\label{NrrC}
\tilde N_{\rm coll}(\vec r_1, \vec r_2)
\es
\int d^2b 
\left[ 1 - e^{- C_2 \tilde N_{\rm coll}(\vec b, \vec r_1, \vec r_2)} \right]  
\nt 
\tilde N_{\rm coll} (\vec b, \vec r_1, \vec r_2) \ , \\
\label{epsrrC}
\tilde \varepsilon_2(\vec r_1, \vec r_2)
\es
\int d^2b 
\left[ 1 - e^{-C_2 \tilde N_{\rm coll}(\vec b, \vec r_1, \vec r_2)} \right]  
\nt 
\tilde \varepsilon_2(\vec b, \vec r_1, \vec r_2) \ .
\eeq
The functions of impact vector $\vec b$ result from
integration over the string overlap,
\beq
\label{Nbrr}
&& \tilde N_{\rm coll}(\vec b, \vec r_1, \vec r_2) \nn
\es  \int d^2x_T \   \tilde  n_{\rm coll}(\vec x_T, \vec b , \vec r_1, \vec r_2)  \ , \\
\label{epsbrr}
&& \tilde \varepsilon_2(\vec b, \vec r_1, \vec r_2) \\
\es
{ [ \{ x_T^2 \cos(2 \alpha) \}^2 
+\{ x_T^2 \sin(2 \alpha)  \}^2 ]^{1/2} 
\over \{ x_T^2 \} } \ , \\
\label{tildenxbrrevf2}
&&
\tilde n_{\rm coll}(\vec x_T, \vec b , \vec r_1, \vec r_2) 
\rs
\hat \rho_1 \ \hat \rho_2 
\nt
A_1(\vec x_T - \vec b/2, \vec r_{T1}) \ A_2(\vec x_T + \vec b/2, \vec r_{T2}) \ , \\
\label{tilderho}
\hat \rho
\es  { r+w  \over (r+w)\cos \beta
+ w \sin \beta }  \ .
\eeq
\end{appendix}



\end{document}